# Neutralization of ion beam by electron injection, Part 2: Excitation and propagation of electrostatic solitary waves

C. Lan[1,2] and I. D. Kaganovich[2]

[1] *Institute of Fluid Physics, China Academy of Engineering Physics, Mianyang, 621900, P. R. China.*
[2] *Princeton Plasma Physics Laboratory, Princeton, New Jersey 08543, USA*

**Abstract:** The charge neutralization of an ion beam by electron injection is investigated using a two-dimensional electrostatic particle-in-cell code. The simulation results show that electrostatic solitary waves (ESWs) can be robustly generated in the neutralization process and last for long time (for more than 30 μs); and therefore ESW can strongly affect the neutralization process. The ESWs propagate along the axis of the ion beam and reflect from the beam boundaries. The simulations clearly show that two ESWs can pass through each other with only small changes in amplitude. Partial exchange of trapped electrons in collisions of two ESWs is observed in the simulations and can explain interaction during collisions of two ESWs. Coalescence of two ESWs is also observed.

## 1 Introduction

Neutralization of positive ion beam space-charge by electrons is important in many areas including accelerators [1], ion thrusters [2], heavy ion inertial fusion [3, 4], and ion-based surface engineering [5], etc. Neutralizing electrons can originate from direct injection of electrons [6], beam-induced ionization of residual gas [7, 8], or plasmas generated by discharge in vacuum [9-18]. In a companion paper [19], hereafter referred to as Part 1, by using two-dimensional particle-in-cell (PIC) simulations we have studied the process of cold electron accumulation during the neutralization of an ion beam. We showed that this process is very slow (compared to the electron bounce time in the potential well of the ion beam), and the beam potential can achieve value much lower than the temperature of injected electrons at the end of neutralization process. It was found that the resulting velocity distribution of neutralizing electrons in the ion beam is non-Maxwellian and exhibits anisotropy. Because neutralizing electrons form double-peak velocity distribution in the direction of the beam propagation (along the *x*–axis) due to bounce motion in the potential well of the ion beam, electrons are subject to the two-stream instability that generates plasma waves, biggest of which form ESWs.

A number of previous works studied the collective behaviors of ion-beam interaction with plasmas. In Refs. [7, 8], M. D. Gabovich *et al.* investigated the collective oscillations in a plasma formed by the beam of fast positive ions and their influence on beam transport. In their experiments, both longitudinal high-frequency and low-frequency waves were observed. Low-frequency waves were presumably caused by ion beam interaction with neutralizing ions and high-frequency waves were excited by electrons. It was observed in these experiments that the excitation of large-amplitude oscillations results in heating of neutralizing electrons. In Refs. [20, 21], authors also studied the electron plasma waves excited by the ion beam as well as possible surface waves in ion-beam plasmas. For the case of ion beam neutralization by volume plasmas, a variety of collective interaction processes can be excited by intense ion beam. For instance, I. D. Kaganovich *et al.* in Refs. [15, 18] performed fully-kinetic PIC simulations of an ion beam pulse entering a plasma, where they observed complex collective phenomena during ion beam entry and exit from the plasma and the excitation of large-amplitude whistler waves in the presence of an applied solenoidal magnetic field. In Refs. [22, 23], the electromagnetic Weibel and electrostatic two-stream instabilities were investigated analytically and numerically for an intense ion beam propagating through background plasma. However, to the best of our knowledge, the possibility of robust excitation of large long-lasting electrostatic solitary waves (ESWs) during neutralization of ion beam has not been investigated or even foreseen in the past decades until our recent work.

ESWs were originally discovered in one-dimensional (1D) two-stream instability simulations [24], and generally they are considered to be a type of Bernstein-Greene-Kruskal (BGK) mode [25]. In the past several decades, they have been extensively studied theoretically [26-28]. In addition, they have been observed in space plasmas and laboratory plasmas for many years [29-38] and were also often referred to by other terms such as electron holes in phase space [39, 40]. In our recent paper [41], we reported that when an ion beam pulse passes through an electron-emitting filament, the generation of ESWs is possible due to the two-stream instability of neutralizing electrons. It was observed in 2D PIC simulations that only one ESW survives for long time with the lifetime of about several microseconds and longitudinal size of about 5 centimeters for simulation parameters corresponding to the Princeton Advanced Test stand experiment [10]. The generation of ESWs has a great impact on the degree of neutralization of ion beam pulse. Possible excitation of ESWs can provide an explanation why past experimental studies [9] showed poorer ion beam neutralization by filaments compared with neutralization by plasmas. However, for neutralization process of a long pulse ion beam (beam length is longer than the chamber length), question whether and which ESWs can be excited requires further study.

In this part of our two-part work, the processes of excitation and propagation of ESWs during the accumulation of cold electrons in a stationary ion beam was investigated making use of a specially written two-dimensional implicit particle-in-cell (PIC) code.

The paper is organized as follows. A 2D simulation model of ion beam neutralization by electron injection is described in the second section of this paper. In section 3, we will show how ESWs are naturally formed during the neutralization of the ion beam in simulations. The mechanism responsible for their excitation will be discussed and the characteristics of ESWs will also be presented. In contrast to the case of short ion beam pulse, several ESWs can be excited in the case of injection of continuous ion beam. Moreover, the lifetime of these ESWs is much longer than those inside a short ion beam pulse. We will also study the collisions of two ESWs in greater details, because of the convenience of producing and controlling ESWs in continuous ion beam.

**2 Simulation model**

The schematics of 2D model used to simulate the ion beam transport in a metal pipe is shown in Fig. 1. Electrons are injected on the axis to neutralize the ion beam. Ion beam, electron injection and transporting metal pipe comprise a simple but complete physical model of neutralization. Such an electron injection scheme represents the electron emission from hot filaments placed across the beam path [6]. For simplicity, the model is 2D in *x-y* uniform Cartesian coordinate system, where *x* is the direction of the ion beam propagation and *y* is the transverse direction. The size of computational domain is 40 cm×6 cm. The cell size of uniform Cartisian grid is 0.25 mm, which leads to a grid of 1600×120 cells. Monoenergetic Ar$^+$ beam with energy $E_b$=38 keV and initial density $n_b = 1.75 \times 10^{14} \text{m}^{-3}$ is injected from the left boundary.

The parameters of ion beam we chose are close to those of Princeton Advanced Test Stand at PPPL [10]. Because ion beam current is very low and ion beam flow velocity $V_b$ satisfies $V_b \ll c$, where *c* is the light speed, the inductive magnetic field of ion beam in vacuum can be neglected compared to its self-electric field, and the system can be treated electrostatically. Therefore, in 2D PIC simulations, Poisson's equation was applied to obtain the electric potential from charge density.

Upper and lower metal walls are totally absorbing boundaries for particles. For left and right boundaries, considering that in experiments ion beams are usually extracted through a metal grid, and collected by a Faraday cup or directly hit a metal target after traveling a distance, so both left and right walls of the model can be treated as metal boundaries for electric field and absorbing boundaries for particles. When ions hit a metal wall, a great number of secondary electrons will be created. However, for simplicity ion induced secondary electron emission was not considered in our simulations.

We also neglected the expansion of the ion beam due to self-space charge, because of very small effect on ESW; expansion of ion beam was considered in the Part 1 paper. Electrons are injected in the center of the domain, i.e., at *x*=20 cm. Electron temperature is 0.2 eV, which is close to the typical temperature of electrons emitted by a hot filament [6, 9]. Electron injection started at *t*=0.8 μs and stopped at *t*=2.4 μs. The value of injected electron current was chosen to be twice of the ion beam current to accelerate the process of neutralization.

The collisions between charged particles and neutral particles were not modeled and coulomb collisions between charged particles were also neglected as they only weakly affect the neutralization process. The time step of the simulations was 80 ps, and single simulation lasted for more than 30 μs. 6000 particles per cell were used to reduce numerical noise. Simulations were run with 40 cores on the Princeton

University Adroit supercomputer.

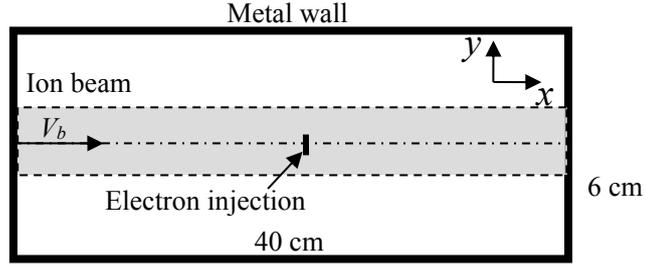

Fig. 1. Schematic of the simulation model. The model is 2D in *x-y* Cartesian coordinates system. Ion beam moves along the axis of the model with speed $V_b$. Electrons are injected on the axis. Dashed curves represent the envelope of ion beam. For simplicity, the expansion of the ion beam is neglected.

**3 Results and discussion**
**A. Excitation of ESWs during the neutralization of the ion beam**

The formation of double-peak $EV_xDF$ leads to the two-stream instability of cold electrons, and the generation of ESWs after this instability evolves into nonlinear stage. In fact, the fluctuations of the beam potential generated at the second stage of ion beam neutralization (see Figs. 3, 7 and 10 in the Part 1 paper) are for this reason. According to the results presented in the Part 1 paper, the neutralization of the ion beam exhibits two distinct stages. At the first stage, all injected electrons are captured by the ion beam, provided that initial beam self-potential is much higher than electron temperature. At the second stage, with the decline of beam potential, hot electrons escape from the ion beam, leaving cold electrons inside the potential well of the ion beam. The second stage lasts for a much longer time than the first stage, providing sufficient time for ESWs to be excited and propagate.

Fig. 2(a) plots the temporal evolutions of potential profile along the *x* axis and Fig. 2(b) shows the corresponding behavior of neutralizing electrons in the $x$-$v_x$ phase space. As evident from this figure ESWs are created during neutralization. Because of the potential drop caused by the space charge near the electron injection position (*x*=20 cm), injected electrons are accelerated first into the ion beam. After reaching near the left and right walls, they are bounced back, generating the two-stream instability of electrons in the phase space. As a result, a lot of small electron holes are created during phase mixing. These small electron holes then coalesce and finally form two big ones near the electron injection position. Once the two big holes get stable, they begin to propagate towards the walls with a velocity on the order of 10 cm/μs, forming two stable ESWs.

The ESWs bounce back and forth between the electron injection point and walls. If the electron injection is stopped, the ESWs can go across the electron injection point and propagate freely between two walls. The reason why the ESWs can be reflected is probably that there is a sufficiently large potential drop near reflection points. The behavior of the ESWs in electric field is similar to particles with the same charge to mass ratio as electron [29, 40]. After a long time of propagation, their amplitudes will gradually decrease and finally decay to zero as the velocity distribution function of neutralizing electrons tends to be Maxwellian. According to Eq. 2 of the Part 1 paper, the duration of the first stage of neutralization is in microseconds. But the lifetime of the ESWs can last for tens of microseconds, much longer than the time required for the first stage of neutralization.

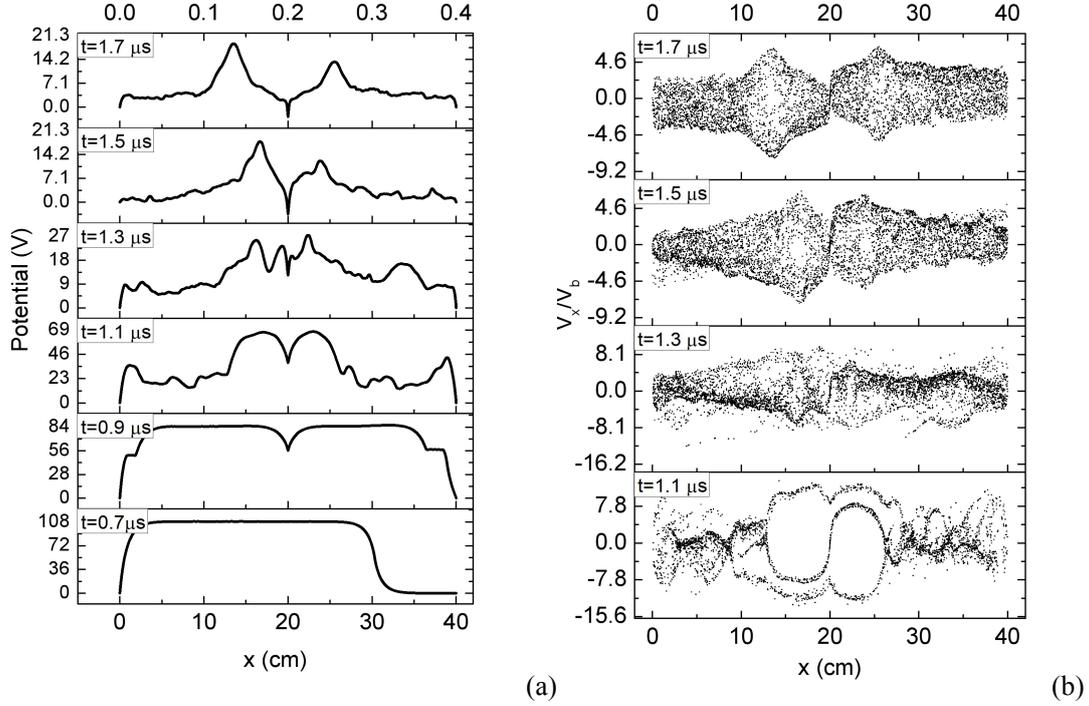

Fig. 2. Temporal evolutions of the beam potential on the axis (a) and electrons in the $x$-$v_x$ phase space (b). The size of the domain in the $y$ direction is 3 cm in this simulation.

Figure 3 plots the profiles of particle densities, potential and electric field along the axis of the ion beam, from which some basic parameters of the ESWs can be obtained. The longitudinal size of the ESWs reaches about 5 cm, which is dozens of Debye lengths. The maximal depth of electron density deficit is about 1/3 of averaged electron density. The positive potential peak caused by localized density deficit of electron can reach more than 10 V. Single humped potential in multi-dimension indicates that along the $x$ direction the parallel electric field is bipolar while the perpendicular electric field is unipolar. This feature can be used to identify ESWs in future experimental studies.

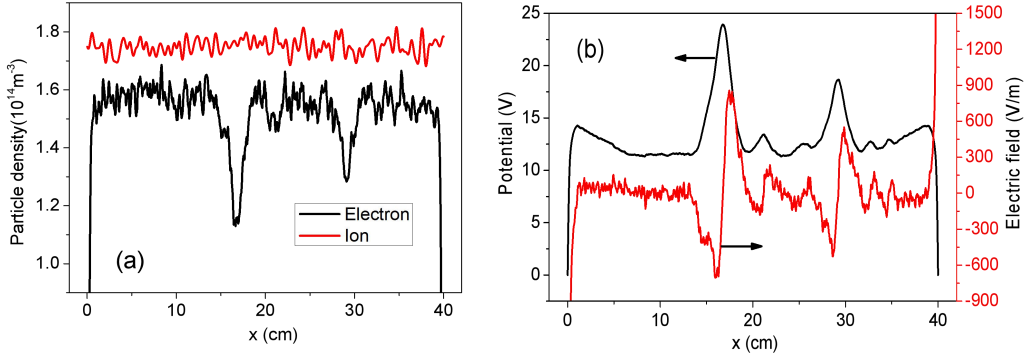

Fig. 3. Profiles of particle densities (a), beam potential and electric field in the *x* direction (b) at *t*=4.2 μs. Note that electron injection is stopped at t=2.4 μs, because of the loss of partial electrons on the walls, electron density is lower than ion density.

In order to obtain the lifetime of the ESWs, we set 5 voltage probes in the simulation to record time-depend beam voltages at these positions. Figure 4 shows plots of potential at different positions as a function of time. Because electron injection ends at 2.4 μs, averaged beam potential after 2.4 μs is much higher than the temperature of injected electrons. Each peak of voltage after 2.4 μs represents an ESW passing through a recording point. Evidently, the peak of voltage at each position gradually disappears after about 35 μs, indicating that the lifetime of the ESWs can last for more than 30 μs. In Ref. [41], we reported that in the case of ion beam pulse, the lifetime of ESW typically is about 4 microseconds. But here, the ESWs are apparently much more stable. Their lifetime of the ESWs is at least 7 times higher than in the previous case. The reason for such long time of duration of the ESWs in 2D is still unclear. It probably attributes to infrequent reflections at both ends of the ion beam, compared with the case of ion beam pulse.

On the other hand, it is clearly seen from Fig. 4 that the attenuation of the ESWs is nearly linear until t≈30 μs. Meanwhile, the averaged beam potential increases very slowly with the attenuation of the ESWs until they disappear at t≈35 μs. After that, the averaged beam potential remains nearly constant (~15 V). The increase of averaged beam potential is related to wave-particle interaction during the propagation of the ESWs that leads to heating of electrons. The most energetic electrons can escape to the walls, which results in reduction of electron population and the increase of beam potential.

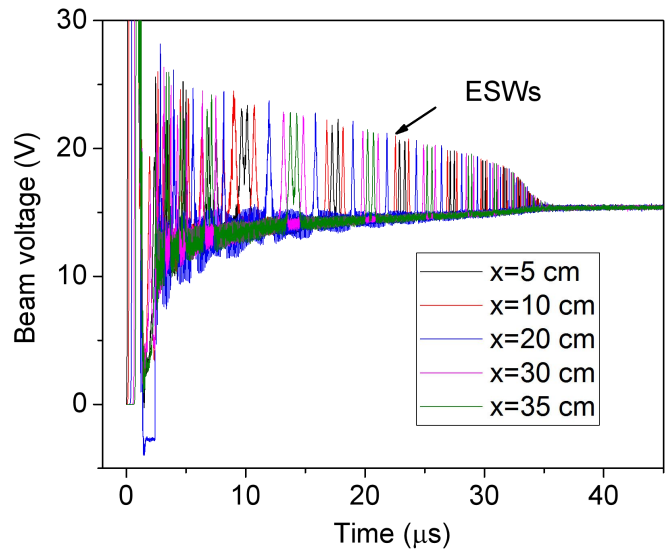

Fig. 4. Temporal evolution of amplitudes of the ESWs. The potentials were recorded at different positions on the axis. When an ESW passes through each recording point, a positive voltage pulse will be recorded.

Details of the motion of the ESWs on an *x-t* diagram are given in Fig. 5. It can be seen that only one ESW can survive after multiple collisions (Collision of ESWs will be discussed below). Then this ESW bounces between the walls with a small acceleration. The acceleration of the ESW may be related to the variation of the velocity distribution of neutralizing electrons. As time goes by, the ESW moves faster and faster. Meanwhile, the amplitude of the ESW is gradually decreased due to damping. It should be noted that in Fig. 5 the time interval between two records is 250 ns. The longitudinal size of the ESW is about 5 cm. Therefore, once the speed of the ESW exceeds 20 cm/μs, the trajectory of the ESW on the *x-t* diagram becomes discontinuous.

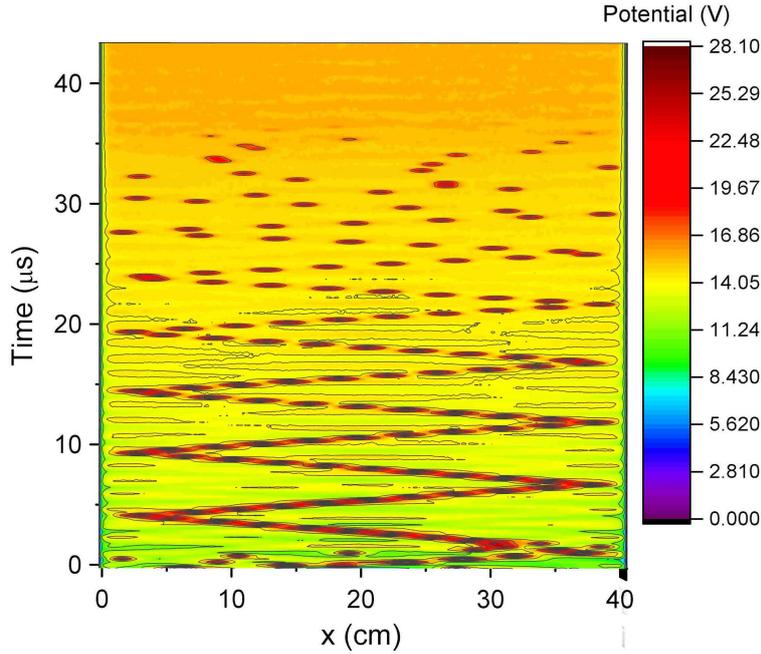

Fig. 5. Spatio-temporal evolution of the trajectories of the ESWs. The dash-like trajectory appeared after about 15 μs is due to relatively large time interval between two records (250 ns) relative to the speed of the ESW.

The excitation of the ESWs occurs due to nonlinear distortions of the electron distribution function during the accumulation of cold electrons. They can only be excited when injected electron current is sufficiently high. In Ref. [42], Omura *et al.* numerically studied the mechanisms responsible for the generation of ESWs. Four instabilities that give rise to ESWs have been studied. The double-peak electron velocity distribution observed in our simulations (see Fig. 6 in the Part 1 paper) more probably leads to the warm two-stream instability [42, 43]. In Schamel's theoretical paper [26], he concluded that for the existence of ESWs, electron velocity distribution must have a concave shape in the trapped electron region. This is clearly presented in Fig. 2(b) and Figs. 7-8 below. Therefore, the generation mechanism and basic characteristics of the ESWs studied here are consistent with previous studies. In contrast to previous simulations about ESWs [24, 38, 42], the ESWs generated during ion beam neutralization are naturally formed and do not require any special setting of initial electron velocity distribution.

Because of electron density depletion in ESW, the excitation of ESWs deteriorates the neutralization of the ion beam. Electron hollows eventually will be filled up with the decay of the ESWs. This process is accompanied by transformation of the electron distribution function towards a Maxwellian distribution.

### B. Propagation and collision of ESWs

As shown in Fig. 2(a), electron injection in the middle of the domain produces a big ESW on each side. Their amplitudes can be controlled by adjusting the position and timing of electron injection. When electrons are no longer injected, these two

ESWs collide with each other. Therefore, current setup is convenient for studying the collision of two solitary waves.

Figure 6 shows plots of temporal evolutions of potential profile along the *x*-axis when two ESWs collide. Two collision events were observed in our simulations. Figure 6(a) shows two ESWs pass through each other, with their identity almost preserved. When their positions coincide, their potential amplitudes are superimposed (not shown in Fig. 6(a)). However, both ESWs experience an abrupt acceleration during the collision. After the collision, two ESWs still move forward, but both of them slightly slow down compared to the original speed. Figure 6(b) shows how these two ESWs collide again and coalesce into a larger ESW after about 2 μs. They first pass through each other in part, with abrupt acceleration towards each other. Then the small ESW moves in the opposite direction under mutual attraction and finally merge into the larger soliton. The resulting solitary wave has similar amplitude but becomes wider in longitudinal size. It moves in the same direction as the larger one before the collision, but at a much slower speed (~3.8 cm/μs).

The abrupt acceleration of two ESWs probably indicates that trapped electrons in ESW's potential well are exchanged in this collision. The relative speed between ESWs is a possible factor that determines whether they pass through each other or coalesce [26, 29]. This is confirmed in our simulations. In Fig. 6(a), the relative speed of two ESWs before collision is about 59 cm/μs, while in Fig. 6(b), this value is only 38 cm/μs. In our simulations, we observed that as time goes by, the motion of ESW is slow down, due to the variation of EVDF during the accumulation of cold electrons (see Figs. 6 in the Part 1 paper). Therefore, it is possible for us to observe two collision events of ESWs at different relative speeds.

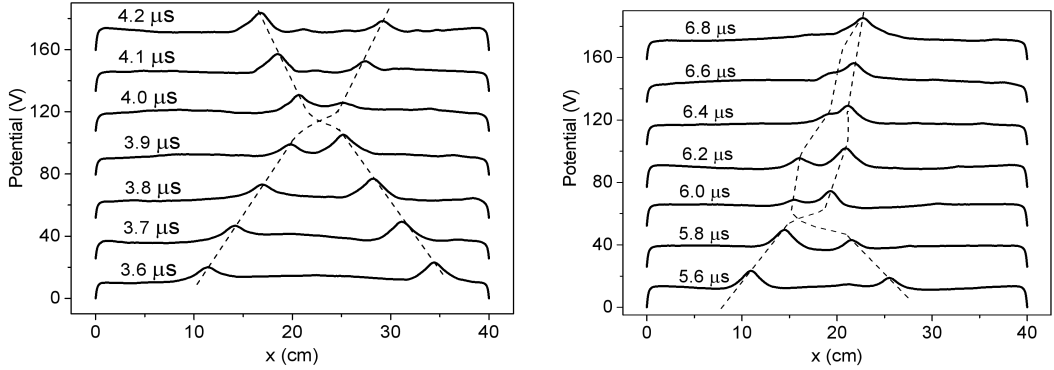

Fig. 6. Collision of two ESWs: (a) passing through, (b) coalescence. Dashed curves represent soliton trajectories on a space-time diagram.

If their relative speed is sufficiently large, two colliding ESWs will be completely staggered in the vertical direction of the phase space. As a result, their trapped electrons will not be exchanged during collision and two ESWs will pass through each other without coalescence [42]. However, the collision of two solitary waves shown here is obviously not this case. Figure 7 shows corresponding behavior of electrons in the $x$-$v_x$ phase space when two ESWs collide and pass through each other. We see two

ESWs are completely overlapped in the vertical direction of the phase space, so it is impossible for them to go around each other without any interference.

As evident from Fig.7 showing details of evolution of the electron phase space, the ESW collision process is very complicated, accompanying by rotation and distortion of electron holes. This is one of many typical collision processes, and we assume that the recording of collision starts at some time $t_0$. At this moment, two counter propagating ESWs are very closed and about to collide. Because the small electron hole travels in the positive direction while the big one travels in the negative direction, under abrupt acceleration the small electron hole moves to the top of the big one when they approach and collide, with a very thin electron layer between them. This thin electron layer almost completely separates trapped electrons of two colliding electron holes during the collision. We believe the existence of phase space electron layer is a necessary condition for preserving their identity after collision. But under what condition the thin electron layer would appear or disappear is still unknown. After reaching the top of the big electron hole, the small electron hole begins to rotate, until they are completely separated in the space. But the rotation of the big electron hole is not obvious. During the rotation and separation of the small electron hole, it can be seen that the thin electron layer disappears, resulting in partial mixing of trapped electrons of two ESWs. This indicates that at least parts of trapped electrons are redistributed between them during the collision.

In order to preserve the identity of ESWs, collision process must ensure that the distribution of trapped electrons in the phase space is unchanged before and after collision. However, the partial mixing of trapped electrons makes this hard to achieve. Consequently, the amplitude of the small electron hole is not preserved and becomes smaller after the collision, as shown in Fig. 7.

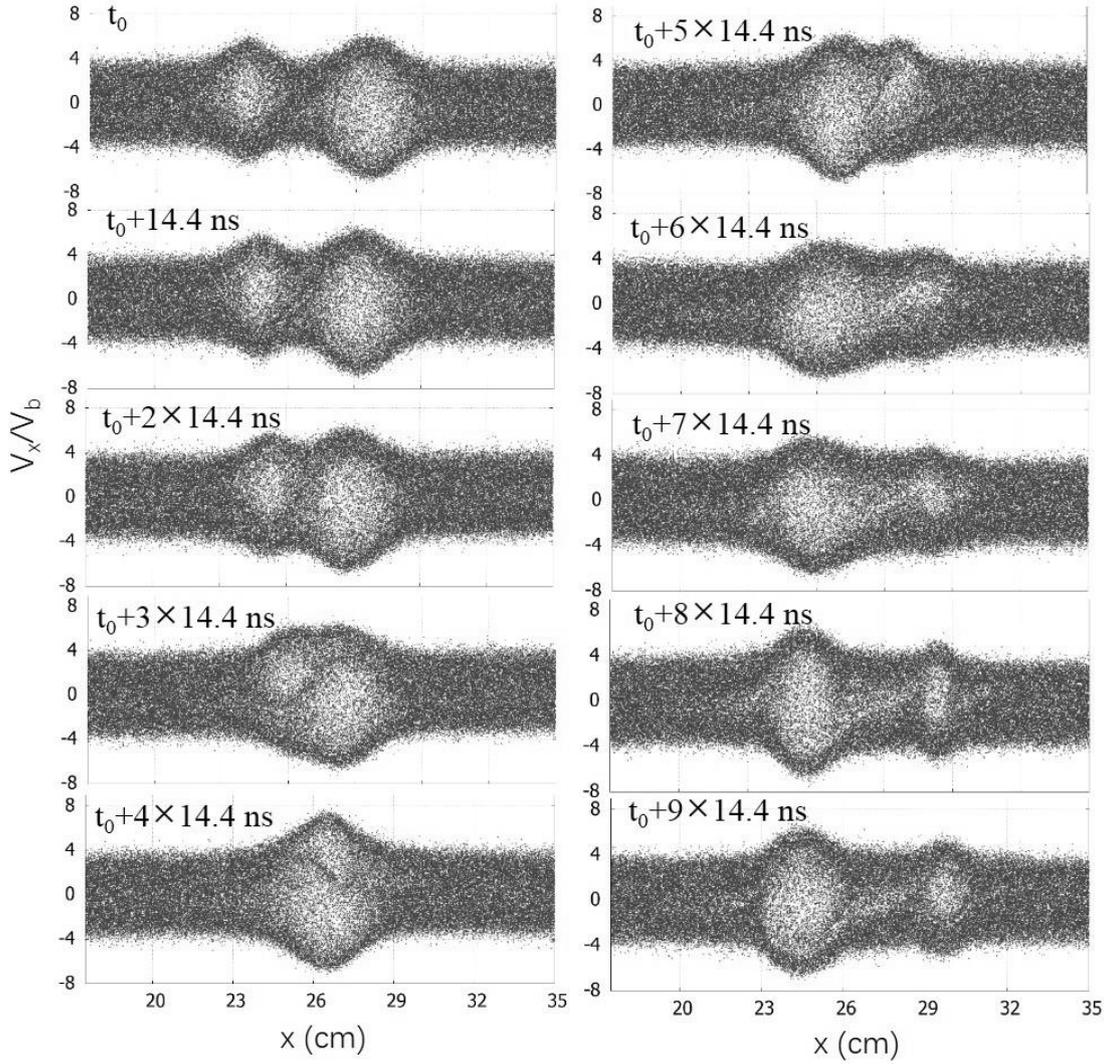

Fig. 7. Evolution of electrons in the *x*-$v_x$ phase space when two ESWs passing through each other. $t_0$ only represents an initial recording time. At this moment, two ESWs are about to collide.

When two ESWs merge into one ESW, typical evolution of the electron phase space is shown in Fig. 8. Similarly, in the first half of the collision two electron holes bypass each other in the phase space. But their trapped electrons are not completely separated. Because their relative speed is not high enough, two partially separated electron holes are unable to escape from their common potential well. As a result, in the second half of the collision, their trapped electrons move reversely and merge together. The resulting electron hole becomes wider in the longitudinal direction, so as to accommodate more trapped electrons.

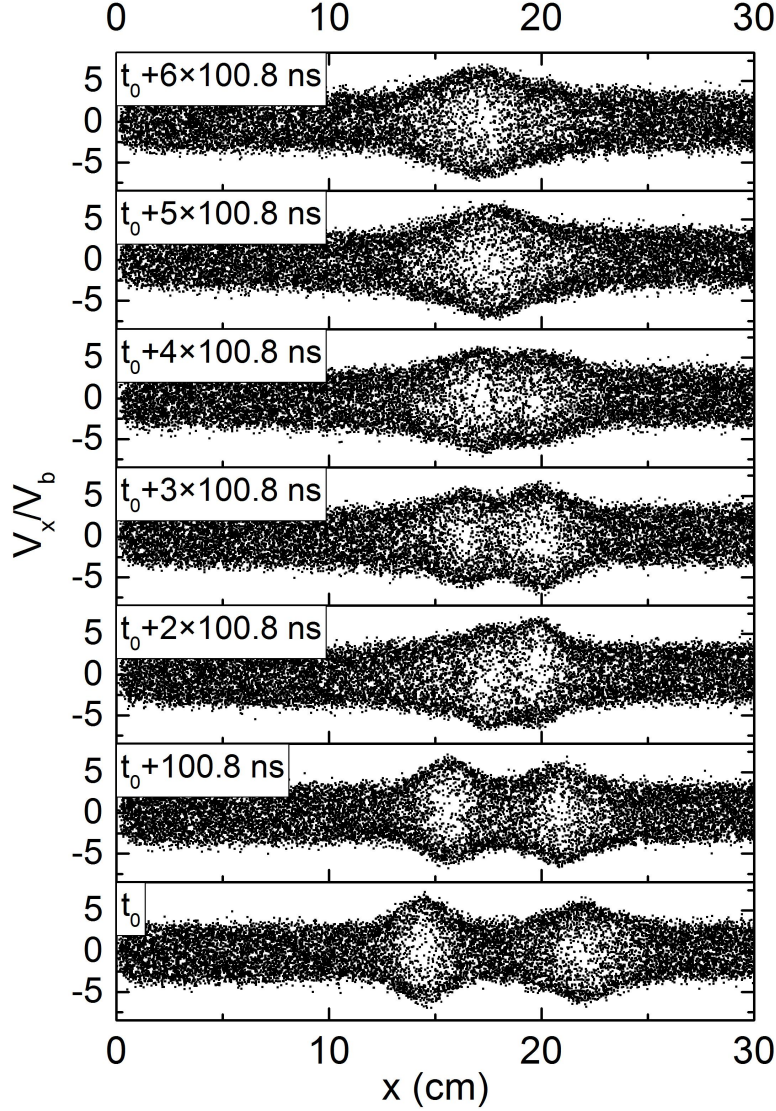

Fig. 8. Evolution of electrons in the *x*-*v_x* phase space when two ESWs are merging into one. $t_0$ only represents an initial recording time. At this moment, two ESWs are about to collide.

**4 Conclusions**

In this paper, we have presented 2D particle-in-cell numerical studies of excitation and propagation of ESWs during neutralization of an ion beam by electron injection. Our simulations show that due to the two-stream instability of neutralizing electrons, the ESWs are excited near electron injection location. The ESWs are formed due to nonlinear distortions of the electron distribution function, which are large enough to produce robust long-lived ESWs that move along the ion beam propagation direction and are bounced back when they reach high electric field areas at the walls or at electron injection location. Because electron density is depleted in the ESWs, their presence strongly affects the degree of neutralization.

When two ESWs collide with each other, they can either pass through each other or coalescence. Surprisingly in both cases ESWs are completely overlapped in the vertical direction of the electron phase space as observed in the simulations. Temporal evolutions of the electron phase space clearly show the whole processes of two ESWs passing through each other and merging into one ESW. Because trapped electrons of two ESWs are partially interchanged in collisions, the ESW's amplitudes are not conserved after they pass through each other. If the relative speed of two ESWs is sufficiently large, they become completely separated after their acceleration towards each other during collisions.

Acknowledgments: The work of Igor D. Kaganovich was supported by the U.S. Department of Energy. The work of Chaohui Lan was supported by the International Cooperation and Exchange Fund of China Academy of Engineering Physics.